\documentclass[manuscript]{aastex}
\usepackage[]{natbib}
\usepackage{graphics}
\usepackage{amssymb}
\usepackage{amsmath}

\slugcomment{} \shorttitle{0846+51W1} \shortauthors{ZHOU ET AL.}

\begin{document}

\title{Hybrid nature of 0846+51W1: a BL Lac object with a narrow line Seyfert 1 nucleus}
\author{Hong-Yan Zhou, Ting-Gui Wang, Xiao-Bo Dong, Cheng Li, and Xue-Guang Zhang }
\email{mtzhou@ustc.edu.cn}

\affil{Center for Astrophysics, University of Science and Technology of
China, Hefei, 230026, China}

\begin{abstract}
We found a NLS1 nucleus in the extensively studied eruptive BL
Lac, 0846+51W1, out of a large sample of NLS1 compiled from the
spectroscopic dataset of SDSS DR1. Its optical spectrum can be
well decomposed into three components, a power law component from
the relativistic jet, a stellar component from the host galaxy,
and a component from a typical NLS1 nucleus. The emission line
properties of 0846+51W1, $FWHM(H\beta)\simeq 1710~ km~ s^{-1} $
and $\frac{[OIII]\lambda 5007}{H\beta}\simeq 0.32$ according to
its SDSS spectrum observed when it was in faint state, fulfil the
conventional definition of NLS1. Strong FeII emission is detected
in the SDSS spectrum, which is also typical of NLS1s. We try to
estimate its central black hole mass using various techniques and
find that 0846+51W1 is very likely emitting at $a~few \times 10\%$
of Eddington luminosity. We speculate that Seyfert-like nuclei,
including NLS1s, might be concealed in a significant fraction of
BL Lacs but have not been sufficiently explored due to the fact
that, by definition, the optical-UV continuum of such kind of
objects are often overwhelmed by the synchrotron emission.
\end{abstract}


\section{INTRODUCTION}
In 1985, Osterbrock \& Pogge identified a special class of active
galactic nuclei (AGN) denominated as Narrow Line Seyfert 1
galaxies (NLS1s), which are characterized by their narrow Balmer
emission line width of $FWHM(H\beta)\lesssim 2000~km~s^{-1}$ and
forbidden to permitted line ratio $\frac{[OIII]\lambda
5007}{H\beta}\lesssim \frac{1}{3}$ (for the conventional
definition of NLS1s, see Pogge 2000). Succedent studies reveal
their extreme properties, such as the narrowest $H\beta$ line
width, the weakest $[OIII]\lambda 5007$, the strongest FeII
emission and the steepest soft X-ray slope. These characteristics
locate NLS1s at the extreme end of the Eigenvector 1 (E1)
parameter space (Boroson \& Green 1992; Sulentic et al. 2000) and
their unusualness is very useful to test the viabilities of AGN
models.

It is found that radio-loud AGN occupy a restrict E1 parameter
space opposite to NLS1s which are usually radio-quiet (Sulentic et
al. 2003). Only three radio-loud NLS1s with radio-loudness
(defined as $\frac{f_{radio}}{f_{optical}}$) between $\sim 10-100$
are known before 2000. They are all radio-intermediate sources
according to the criterion of Sulentic et al (2003). About twenty
more radio-loud NLS1s were identified by Whalen et al. (2001) in
the FIRST Bright Quasar Survey. Zhou \& Wang (2002) found eight
more radio-loud NLS1s by cross-correlating the Veron \&
Veron-Cetty AGN with the FIRST and NVSS radio catalog. Again,
almost all of these newly dig out radio-loud NLS1s are moderate
radio sources. It is remarkable that all of the discovered
radio-loud NLS1s are compact at the present spatial resolution,
indicating that relativistic beaming might be important in at
least some of these objects. SDSS J0948+0022 is hitherto
the only genuinely very radio-loud NLS1 possibly with a
relativistic jet beaming toward the observer (Zhou et al. 2003).
These properties call to remembrance of blazar, which is another
small divertive subset of AGN.

Blazar is a pictorial term first proposed by Spiegel in 1978 be
applied to rapidly variable objects (see Burbidge \& Hewitt 1992)
and all of the known blazars are radio sources. Nowadays it is
believed that blazars, including OVVs (optically violent
variables) and BL Lacs, are those AGN that have a strong
relativistically beamed component close to the line of sight. BL
Lacs and OVVs share many common properties except that, by
definition, in the former emission line is very weak or absent.
However, the rarity of very radio-loud NLS1s, whose occurrence in
low redshift broad line AGN is estimated $\lesssim 0.2\% $ (Zhou
et al. 2003) is in the way addressing ourselves to this problem.

The large sky area coverage and moderate deepness of the Sloan
Digital Sky Survey (SDSS\footnote{The SDSS Web site is
http://www.sdss.org/.}, York et al. 2000) make it be propitious to
exploring rare objects such as very radio-loud NLS1s. In this
letter we report the discovery of another such object,
0846+51W1=SDSS J084957.98+510829.1, out of $\sim 500$ NLS1s
compiled from the spectroscopic dataset of the SDSS Data Release 1
(DR1). It is even more conspicuous than SDSS J094857.3+002225 and
shows many dramatic properties.

Actually 0846+51W1 was originally found by Arp et al. (1979) and
has been the subject of much study since then. This object is
violently variable in its optical flux ($\Delta V \sim 5^{m}$ over
a time span of $\sim 1$ year and $\Delta V \sim 4^{m}$, from
$V\simeq 15^{m}.8$ to $V\sim 19^{m}.5$ within one month). At its
maximum light burst, the optical spectrum was found to be
featureless, while emission lines were detected when it became
fainter. Its optical slope may vary dramatically from
$\alpha_{o}\approx 1.6$ when the object is bright to
$\alpha_{o}\approx 2.8$ when it is faint ($f_{\nu}\varpropto
\nu^{-\alpha_{o}}$, c.f. Arp et al. 1979; Stickel et al. 1989). It
is also found to be highly polarized in both of the radio and
optical bandpass (Moore \& Stockman 1981; Sitko et al. 1984).
Therefore 0846+51W1 bears all the characteristics of BL Lac
object. However, the narrow wavelength coverage of Arp et al. and
Stickel et al. led 0846+51W1 be taken as a high redshift BL Lacs
(z=1.86) by these authors whereas the SDSS spectrum clearly shows
that its true redshift is $z=0.5835$.

We will analysis the SDSS optical spectrum of 0846+51W1 in detail
in \S 2. Some implication of 0846+51W1 is discussed in \S 3. The
main purpose of this letter is to reignite further interest to
this fascinating object. We adopt a $\Lambda$-dominated cosmology
with $H_0=70~km~ s^{-1}~Mpc^{-1}$, $\Omega_M=$ 0.3 and
$\Omega_\Lambda=$ 0.7 through out this letter.

\section{OPTICAL SPECTRUM ANALYSIS}

0846+51W1 is observed by SDSS only because it is the counterpart
of a FIRST radio source and its SDSS spectrum is shown in Figure 1
(denoted as thin line) with recognizable emission lines labelled.
At first sight, the optical spectrum of 0846+51W1 looks like a
typical NLS1 and by no means let any association with a BL
Lacertae occur to us though the spectrum is rather noisy. Almost
all of the familiar emission lines, even the weak ones such as
[NeV]$\lambda 3346$, [NeV]$\lambda 3426$, [NeIII]$\lambda 3869$
and [NeIII]$\lambda 3888$ are present in the spectrum. Optical and
UV FeII multiplets can also be easily spotted out. The spectrum
were taken at 2002 November 29 and the visual magnitude estimated
from the spectrum is about $19^{m}.8$ indicating that 0846+51W1
was in its quiescent state when it was observed. At a redshift of
$z=0.5835$, its luminosity is $M_{V}\simeq -22^{m}.1$, which is
comparable to luminous CD galaxies indicating that the host may
contribute significantly to the observed spectrum. Many authors
invoke gravitational (micro)lenses to interpret the eruptive
behavior of 0846+51W1 (e.g., Nottale 1986; Stickel et al. 1989).
However, the HST image when 0846+51W1 was in its faint state
($V\simeq 19^{m}.7$) did not show significant resolved structure
(Bahcall et al. 1993) and, with broader wave coverage and higher
resolution, the SDSS spectrum does not show the signature of any
intervening galaxy. Because the S/N ratio of the SDSS spectrum is
not high, we model it with three components
\begin{equation}\label{eq1}
f(\lambda)=aA(\lambda)+bB(\lambda)+cC(\lambda)
\end{equation}
where $A(\lambda)$ denotes a composite spectrum of NLS1 (cf.
Constantin \& Shields 2003), $B(\lambda)\varpropto \lambda^{-0.5}$
is a power law component from the relativistic jets, $C(\lambda)$
is the template of elliptical galaxy (cf. Mannucci et al. 2001)
and a, b, and c represent the relative contribution of the three
components which are set free. Emission lines except FeII
multiplets are masked in the fitting procedure. The final fit is
done through minimization of $\chi^{2}$ and the result is
acceptable (reduced $\chi^{2}=1.36$) and is also shown in Figure
1.

Now the continuum subtracted spectrum is fitted to measure
prominent emission line parameters. The $H\beta +[OIII]\lambda
\lambda 4959,5007$ regime is fitted with one Lorentzian + three
Gaussians. The $H\beta $ is modelled by one Lorentzian (broad
component) and one Gaussian (narrow component) and the
$[OIII]\lambda \lambda 4959,5007 $ doublet are fitted with two
Gaussians. The width and redshift of the three Gaussians are
forced to be the same and the intensity ratio of the
$[OIII]\lambda \lambda 4959,5007 $ doublet is fixed to the
theoretical value. We also force the ratio of $[OIII]\lambda 5007$
to $H\beta $ narrow component to be equal to 10, which is the
typical value of Seyferts. This is a common procedure to prevent
the fitting routine to yield non-physical value when spectrum is
noisy (e.g., Veron et al. 2001). Almost all known NLS1s show
strong FeII emission as does 0846+51W1. However, measurement of
its FeII multiplets is rather challenged for the present spectral
quality. Judging from the fitting residual of Equation (1), what
we can say is that its optical FeII strength should be at least
comparable to, possibly stronger than that of typical NLS1s. We
did not analyses other emission lines because the quality of the
spectrum is not high enough for us to draw any significant
conclusion. The main parameters of prominent emission lines are
listed in Table 1.

\begin{table*}
\begin{minipage}{180mm}
{\bf Table 1.} Emission line parameters of 0846+51W1. \\[0.1cm]
\nobreak
\begin{tabular}{lccc}
\hline \hline
line & Flux & FWHM \\
 & 10$^{-17}$erg s$^{-1}$ cm$^{-2}$ & $km s^{-1}$ \\
\hline
$[OIII]\lambda 5007$ & 45$\pm $4.7 & 372$\pm $31   \\
$H\beta $(broad component) & 139$\pm $22 & 1710$\pm $184   \\
$MgII\lambda $2800 & 286$\pm $39 & 2512$\pm$471 \\
FeII$\lambda \lambda $4570 & $\gtrsim$ 100 &   \\

\hline
\end{tabular}
\end{minipage}
\end{table*}

\section{DISCUSSION}

\subsection{THE CENTRAL BLACK HOLE MASS AND THE ACCRETION RATE OF 0846+51W1}
Mass estimates for the central black holes in AGN have become
feasible using various techniques and the exact value of $M_{BH}$
of 0846+51W1 is of much interesting considering its dualism of
NLS1 and BL Lac object. We will try to estimate $M_{BH}$ using
three distinct approaches and make comparison among the results
yielded.

We first use the virial assumption with FWHM(H$\beta$) measured
and the decomposed luminosity of the disk component ("A" component
of \S2). Then $M_{BH}$ can be estimated as (Kaspi et al. 2000)
\begin{equation}\label{eq2}
M_{BH}=1.464\times10^{5}(\frac{R_{BLR}}{lt-days})(\frac{v_{FWHM(H\beta)}}{10^{3}km
s^{-1}})^{2}~ M_{\odot}
\end{equation}
where $R_{BLR}$ is the size of broad $H\beta$ emission line region
which is related to the monochromatic luminosity of "A" component
through the empirical relation (Kaspi et al. 2000 converted to our
adopted cosmology)
\begin{equation}\label{eq3}
R_{BLR}=22.3[{\frac{\lambda L_{\lambda}(5100\AA)}{10^{44}erg
s^{-1}}}]^{0.7}~ lt-days
\end{equation}
For the measured value of $S_{\lambda}(5100\AA)\simeq 1.2\times
10^{-17}~ erg~ s~ cm^{-2}~\AA $ ("A" component) and
$FWHM(H\beta)\simeq 1710~km~ s^{-1}$ we obtain $M_{BH}\simeq
8.2\times 10^{6}~M_{\odot}$.

The second method to estimate $M_{BH}$ is to use $[OIII]\lambda
5007$ as a surrogate for the stellar velocity dispersion of the
bulge $\sigma_{*}$ and the $M_{BH}-\sigma_{*}$ correlation
(Tremaine et al. 2002)
\begin{equation}\label{eq4}
log(\frac{M_{BH}}{M_{\odot}})=(8.13\pm0.06)+(4.02\pm0.32)log(\frac{\sigma_{*}}{200~km~s^{-1}})
\end{equation}
If the motion of [OIII] emission line clouds in the NLR of AGN is
dominated by the gravitational potential of the host galaxy bulge,
a strong correlation between FWHM[OIII] and $\sigma_{*}$ would be
expected and such a correlation is indeed found by Nelson \&
Whittle (1996). Nelson (2000) suggested that FWHM[OIII] may be
used as a surrogate for $\sigma_{*}$ by the relation
\begin{equation}\label{eq5}
\sigma_{*}\sim \frac{FWHM[OIII]}{2.35}
\end{equation}
Using Equation (3) and Equation (4) we obtain $M_{BH}\simeq
5.2\times 10^{7}~ M_{\odot}$.

The third way to estimate $M_{BH}$ is to use the
$M_{BH}$-$M_{bulge}$ correlation. The $M_{bulge}$ can be deduced
adopting the $L/M$ ratio of elliptical galaxies where L correspond
to the "C" component of \S2. Using the the following empirical
relation (Laor 2001 converted to our adopted cosmology)
\begin{equation}\label{eq6}
M_{V}(bulge)=-10.06\pm 1.08 - (1.38\pm 0.13)
log(\frac{M_{BH}}{M_{\odot}})
\end{equation}
with the decomposed host luminosity of $M_{V}(bulge)\simeq
-20^{m}.5$, we got $M_{BH}\simeq 4.3\times 10^{7}~ M_{\odot}$.

Considering the large uncertainty of these techniques ($\sim
0.4-0.7$ dex, Vestergaard 2004), the estimated values of $M_{BH}$
are consistent with each other. Because its nonthermal emission is
highly boosted, it is difficult to estimate the bolometric
luminosity of 0846+51W1 and we use the emission of "A" component
to yield its lower limit. Assuming that the bolometric luminosity
is about nine times of the monochrome luminosity at B-band (Elvis
et al. 1994), the Eddington mass $M_{Edd}$ should be $> 5.3\times
10^6 M_{\odot}$. Hence 0846+51W1 should emit at $>10\% $ of
Eddington luminosity.

\subsection{On the hybrid nature of 0846+51W1}
The SDSS spectrum of 0846+51W1 is typical of NLS1s. The width of
$H\beta $, $FWHM(H\beta)\approx 1700 km s^{-1}$ is narrower than
normal broad line AGN and the flux ratio of $[OIII]/H\beta \approx
0.3$ excludes the possibility that it might be a type 2 AGN. The
primary ionization source should be the thermal component from the
accretion disk because the little blue bump clearly presents
itself. However, its soft X-ray photon index $\Gamma_{0.2-2.4
keV}=0.61^{+0.38}_{-0.49}$ is very flat. Such a difference between
0846+51W1 and "normal" NLS1s can be anticipated because in the
X-ray we may be observing mainly the jet emission as in other BL
Lacs.

Considering that NLS1s show extreme characteristics opposite to
classical radio-loud AGN, the nature of radio sources in 0846+51W1
is of particular interesting. Accumulated evidence indicates that
the physical driver of Eigenvector 1 may be the accretion rate of
active nuclei, with source orientation playing a concomitant role
and NLS1s are believed to have large $\dot{m}\equiv
\frac{\dot{M}}{M_{BH}}$ and small inclination angle $i$. It may be
exactly the case of 0846+51W1 and the other very radio-loud NLS1,
SDSS J0948+0022. The radio power of 0846+51W1 $P_{5 GHz}\sim
3.9\times 10^{26} W~ Hz^{-1}$ (radio fluxes are adopted from Arp
et al. 1979 and Gregory \& Condon 1991) is comparable to SDSS
J0948+0022, which is also a highly variable object (Zhou et al.
2003). It has been found that in quasars with no Doppler boosting,
the luminosity of $H\beta $ is tightly correlated with the
continuum luminosity with the median rest-frame $H\beta $
equivalent width $EW_{H\beta} \sim 80 $ \AA~ (Veron-Cetty \& Veron
2000). We note that $EW_{H\beta}\simeq 18 $ \AA~ of 0846+51W1 and
25 \AA~ of SDSS J0948+0022 are both much less than the above
median value. This indicates that synchrotron emission from
relativistic jet might make significant contribution to the
optical continuum. Indeed, we would have $EW_{H\beta}\simeq 73$
\AA for 0846+51W1 if the underling continuum of "A" component is
used in calculation. We argue that 0846+51W1 and SDSS J0948+0022
are actually NLS1s oriented with jet axis almost along our line of
sight and consequently extremely beamed and henceforth show
blazar-like behavior. If the Doppler factor is $\gtrsim 10$, the
intrinsic radio luminosity of these two objects would be around
the FR I/FR II transition.

Arp et al. (1979) found that 0846+51W1 fulfils four of the five
criteria for BL Lac object, i.e., weak line feature, large
amplitude variability, nonthermal continuum and red color, except
that polarization measures were not available then and strongly
argued that 0846+51W1 should be classified as a BL Lac object.
High polarization of $> 10\% $ was also detected in the optical
and radio by Moore \& Stockman (1981) and Sitko et al. (1984).
According to the current unified schemes of AGN, the parent
population of blazar are radio galaxies (RGs) and radio-loud
quasars (RLQs). The radio morphology of RGs and RLQs are
classified to two categories according to Fanaroff \& Riley
(1974): FR IIs are the classical double radio sources with
edge-bright lobes while FR Is have edge-darkened morphologies. BL
Lacs are taken as "beamed" FR I radio galaxies while OVVs as
beamed FR IIs. The main deference between the two is believed to
be the accretion rate, which is rather small in the former. For
$\dot{m}\lesssim 1\%$, the accretion flow is advection dominated
with small radiative efficiency. In this scenario, neither FR I
nor BL Lac can be associated with optically powerful quasar.
However, broad line AGN and high luminous quasar associated with
FR I radio structure have been reported in recent year (Lara et
al. 1999; Blundell \& Rawlings, 2001). Correspondingly, broad
emission lines have also been detected in dozens of BL Lacs and
even in BL Lacertae itself (see Table 3 and 4 of Veron-Cetty \&
Veron 2000).

BL Lacs are conventionally defined as blazars with rest-frame
emission line equivalent widths smaller than 5 \AA~ (Morris et al.
1991). The distribution of emission line equivalent width in
blazar is obviously not bimodal and the strength of the highly
variable synchrotron continuum can further blur such an arbitrary
boundary. Objects that fulfil the above criterion may form a
rather complex family. While the emission line properties of some
BL Lacs mimic LINERs indicating low mass accretion rate, other BL
Lacs including the prototype one may harbor a Seyfert-like nucleus
(e.g., Corbett et al. 2000). Some of the BL Lacs with Seyfert-like
nucleus may have relatively high accretion rate but, in optical-UV
band, the emission lines and thermal continuum from the hot
accretion disc can be overwhelmed by the Doppler-boosted
synchrotron component some of the time. 0846+51W1 is the rare care
at it happens when the synchrotron continuum became faint enough
for the NLS1 nucleus to reveal itself. In accordance with
0846+51W1 which is likely emitting at near Eddington luminosity,
Blundell \& Rawlings (2001) found that a optical powerful quasar
E1821+643 is associated with a FR I radio structure indicating
that relatively high accretion rate can also occur in FR Is.

It is remarkable that, on the one hand, no highly radio-luminous
FR Is have been found as yet, and on the other hand, no NLS1 has
been reported to be very powerful in the radio except SDSS
J0948+0022 and 0846+51W1 which are strongly boosted and their
intrinsic radio luminosity may not be too high. This suggests that
apart from other properties such as the spin rate of the central
black hole, mass accretion rate play an role in determining the
radio power and morphology. When $\dot{m}\ll \dot{m}_{Edd}$, only
low power jet can engender because there is not enough input
energy, which can be easily disrupted and dissipated within short
distance from the core and forms an FR I source (De Young 1993).
Contrarily, under the condition of near or supper Eddington
accretion, neither does the air rich environment lend itself to
the collimation and propagation of the jet, which may be
translated into outflow in some extreme cases. In 0846+51W1 and
SDSS J0948+0022 we are very likely observing the innermost part of
the jet pointing toward us and the fact that all of the radio
sources in NLS1s are compact can be understood according to this
interpretation. We speculate that, while both low and high mass
accretion can occur in FR Is and their beamed cousins BL Lacs, the
accretion rate of FR IIs can only (at one time) be moderately
high. It has been pointed by Blundell \& Rawlings (2001) that
optically luminous FR I quasars are well under-investigated. We
speculate that Seyfert-like nuclei might be concealed in a
significant fraction of BL Lacs but have not been sufficiently
explored due to the fact that, by definition, the optical-UV
continuum of such kind of objects are often overwhelmed by the
synchrotron emission. A few of these objects may be "very"
radio-loud NLS1s provided that they are observed at very small
inclination angle.

\acknowledgments This work was supported by Chinese NSF through
NSF10233030 and NSF19990754, and a key program of Chinese Science
and Technology ministry. This paper has made use of the data from
the SDSS. Funding for the creation and the distribution of the
SDSS Archive has been provided by the Alfred P. Sloan Foundation,
the Participating Institutions, the National Aeronautics and Space
Administration, the National Science Foundation, the U.S.
Department of Energy, the Japanese Monbukagakusho, and the Max
Planck Society. The SDSS is managed by the Astrophysical Research
Consortium (ARC) for the Participating Institutions. The
Participating Institutions are The University of Chicago,
Fermilab, the Institute for Advanced Study, the Japan
Participation Group, The Johns Hopkins University, Los Alamos
National Laboratory, the Max-Planck-Institute for Astronomy
(MPIA), the Max-Planck-Institute for Astrophysics (MPA), New
Mexico State University, Princeton University, the United States
Naval Observatory, and the University of Washington.

\clearpage

\begin{figure}
\epsscale{1.0} \plotone{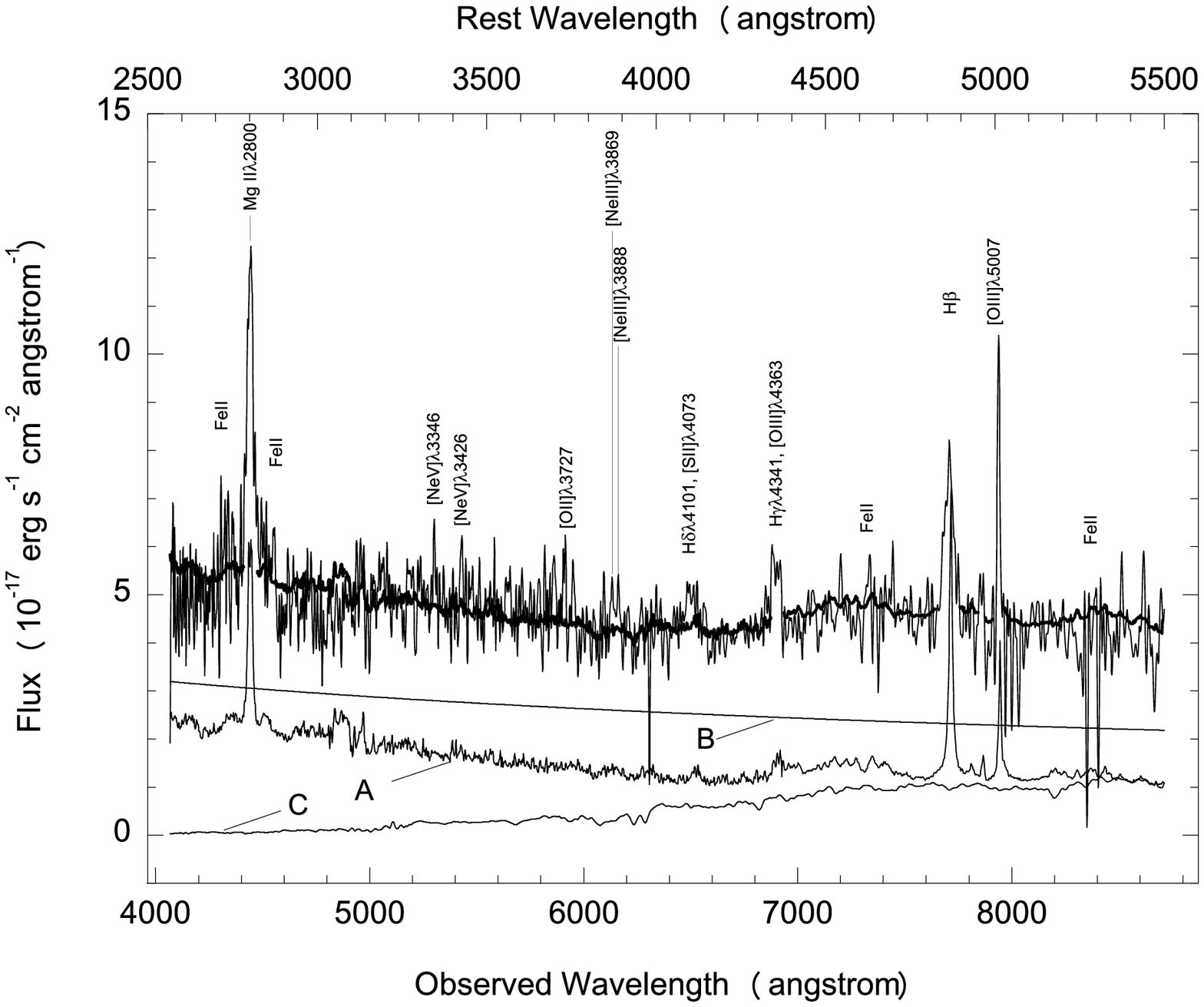} \caption{Optical continuum of
0846+51W1 (the uppermost thin line) can be well decomposed into
three components of A: average NLS1 spectrum, B: spectral index of
$\alpha =1.5$ power law spectrum ($f_{\nu}\varpropto
\nu^{-\alpha}$) and C: elliptical template. The uppermost thick
line is the sum of the three components.} \label{reduction1}
\end{figure}

\begin{figure}
\epsscale{1.0} \plotone{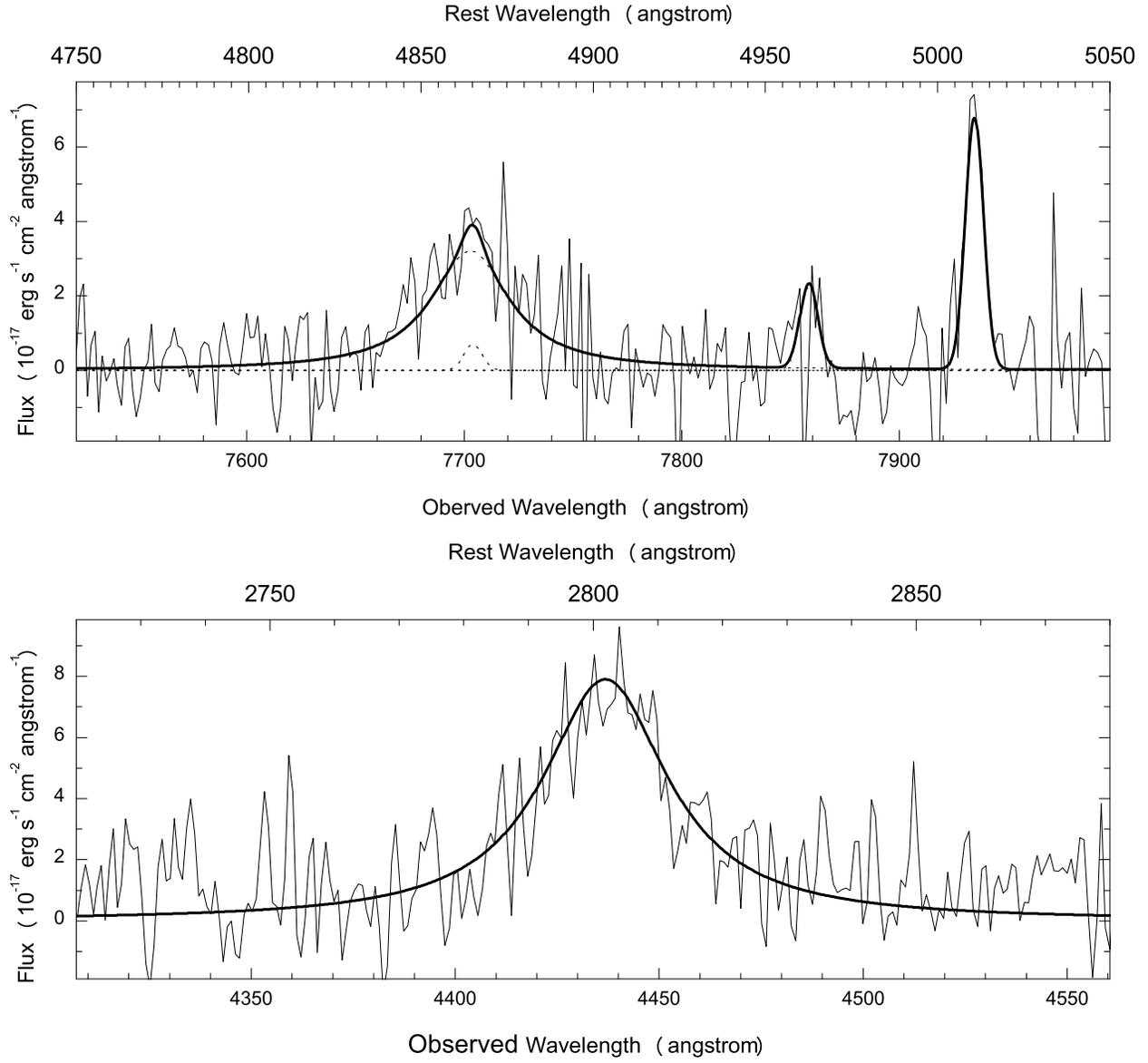} \caption{Both of $H\beta $ and
MgII can be well fitted with Lorentz profile and the line widths
are typical of NLS1s.} \label{reduction2}
\end{figure}

\end{document}